
%
\input phyzzx
\tolerance=1000
\sequentialequations
\def\rl{\rightline}

\def\r#1{$\bf#1$}

\def\t1{{\tilde 1}}

\def\AEF{A.E. Faraggi}

\def\NPB#1#2#3{Nucl. Phys. B {\bf#1} (19#2) #3}
\def\PLB#1#2#3{Phys. Lett. B {\bf#1} (19#2) #3}
\def\PRD#1#2#3{Phys. Rev. D {\bf#1} (19#2) #3}
\def\PRL#1#2#3{Phys. Rev. Lett. {\bf#1} (19#2) #3}

\def\l{\langle}
\def\r{\rangle}
\def\D{$D_{45}~$}
\def\bD{$\bar D_{45}~$}

\REF\GSW{M. Green, J. Schwarz and E. Witten,
Superstring Theory, 2 vols., Cambridge
University Press, 1987.}
\REF\SLM{\AEF, \PLB{274}{92}{47}; \NPB{387}{92}{239}.}
\REF\CAB{A. E. Faraggi and E. Halyo, \PLB{307}{93}{305}.}
\REF\CKM{A. E. Faraggi and E. Halyo, WIS--93/34/MAR--PH.
to appear in Nucl. Phys. B.}
\REF\NB{A. Nelson, \PLB{136}{84}{387}; S. M. Barr, \PRL{53}{84}{329}.}
\REF\DA{F. del Aguila, G. L. Kane and M. Quiros, \PLB{196}{87}{531}.}
\REF\FLQM{S. Kelly, J. L. Lopez and D. V. Nanopoulos, \PLB{261}{91}{424}.}
\REF\MOD{\AEF, \PLB{278}{92}{131}.}
\REF\LQ{E. Halyo, WIS-93/114/DEC-PH, to appear in Phys. Lett. B.}
\REF\FLQ{S. Kelly, J. L. Lopez and D. V. Nanopoulos, \PLB{278}{92}{140}.}
\REF\NRT{\AEF, \NPB{403}{93}{101}.}
\REF\BAR{S. Weinberg, \PRD{26}{82}{287}; N. Sakai and T. Yanagida,
\NPB{267}{82}{533}.}
\REF\UP{E. Halyo, WIS-93/98/OCT-PH, hep-ph 9311300; A. E. Faraggi, IASSNS-HEP
-93/82, hep-ph 9312214, to appear in Phys. Lett. B.}
\REF\PDG{Review of Particle Properties, Phys. Rev. D {\bf 45}, Vol. 11B.}
\REF\NIR{Y. Nir, in ``CP Violation", Lectures presented in the $20^{th}$
Annual SLAC Summer Institute on Particle Physics, 1992 and references therein.}
\REF\DCC{F. del Aguila, M. K. Chase and J. Cortes, \NPB{271}{86}{61}.}
\REF\SNB{M. Dine, R. L. Leigh and A. Kagan, \PRD{48}{93}{2214}.}

\singlespace
\rl{WIS--94/20/APR--PH}
\rl{\today}
\rl{T}
\pagenumber=0
\normalspace
\smallskip
\titlestyle{\bf{CKM Matrix from Leptoquark--Quark Mixing in a String Model}}
\smallskip
\author{Edi Halyo{\footnote\dag{e--mail address: jphalyo@weizmann.bitnet}}}
\smallskip
\centerline {Department of Particle Physics,}
\centerline {Weizmann Institute of Science}
\centerline {Rehovot 76100, Israel}
\vskip 6 cm
\titlestyle{\bf ABSTRACT}

We investigate a scenario in which the CKM matrix arises from leptoquark--down
quark mixing in a particular standard--like superstring model. We find that
for some choices of F and D flat directions realistic quark mixing can be
obtained without any phenomenological problems. This scenario predicts a
symmetric (in absolute value) CKM matrix.

\singlespace
\vskip 0.5cm
\endpage
\normalspace

\centerline{\bf 1. Introduction}

The origin of quark mixing is one of the fundamental questions in particle
physics that the Standard Model does not answer. Extensions of the Standard
Model such as models with supersymmetry or supergravity or grand unified
theories are not improvements in this respect. All we
are able to do with our present
level of knowledge (or ignorance) is to parametrize quark mixing by the
Cabibbo--
Kobayashi--Maskawa (CKM) matrix in terms of three angles. Any theory such as
superstrings [\GSW] which claims to be the fundamental theory must be able to
explain the origin and hopefully the magnitude of quark mixings.
In the framework of standard--like superstring models [\SLM], quark mixing was
investigated in Refs. [\CAB] and [\CKM]. There, it was shown that nonzero
off--diagonal
elements in the up and down quark mass matrices require that some of the states
$V_i,\bar V_i$ from the hidden sectors $b_i+2\gamma$ get VEVs (due to the
generational gauged $U(1)$ symmetries of these models). The source of quark
mixing was identified to be the the VEVs of the hidden sector states
$V_i, \bar V_i$. A specific set of scalar VEVs was shown to give correct
order of magnitude quark mixing angles.

In this letter, we show that quark mixing can also arise as a result of the
the presence of a pair of $TeV$ scale leptoquarks which mix with the left and
right--handed down quarks. Correct order of magnitude mixing angles can be
obtained from proper amounts of leptoquark--down quark mixings which require
specific scalar VEVs around $M_{Pl}$. This scenario also satisfies the
constraints from the unitarity of the CKM matrix and flavor changing neutral
currents (FCNC) due to $Z$ exchange. We also investigate the related issue of
the Nelson--Barr mechanism [\NB] as a solution to the strong CP problem. We
find
that in that case one cannot generate large enough quark mixing (and weak CP
violation).

Similar ideas have been explored before either in a general framework [\DA]
or in flipped $SU(5) \times U(1)$ superstring models [\FLQM] for $SU(2)_L$
singlet, heavy down quarks. In the former no concrete
model was considered and the whole discussion was generic. In the latter, on
the other hand, no estimate of quark mixing was made due to the lack of
calculational tools. In this letter, we consider a specific superstring model
in which reliable estimates of all relevant terms can be made.

For concreteness we consider the generic standard--like superstring model
of Ref. [\MOD] which has leptoquarks in the massless string spectrum.
The complete massless
spectrum with the quantum numbers and the cubic superpotential was presented
in Ref. [\MOD] and will not be reviewed here. The notation of Ref. [\MOD] is
used throughout this letter.

\bigskip
\centerline{\bf 2. Leptoquark--down quark mixing}

In the massless $b_1+b_2+\alpha+\beta+(S)$ sector of the standard--like
superstring model under consideration, there are two color triplet,
electroweak singlet states, $D_{45}$ and $\bar D_{45}$ [\MOD].
Under $SU(3)_C \times SU(2)_L \times U(1)_C \times U(1)_L$, $D_{45}$ and
$\bar D_{45}$ transform as $(3,1,-1,0)$ and $(\bar 3,1,1,0)$ respectively.
In these models, $Q_Y=Q_C/3+Q_L/2$ and $Q_{Z^\prime}=Q_C-Q_L$ and therefore
we find that $Q_Y(D_{45})
=Q_{EM}(D_{45})=-1/3$ and $Q_{Z^\prime}(D_{45})=-1$ with $\bar D_{45}$ having
the opposite charges. On the other hand, $Q_{B-L}=2Q_C/3$
which is a gauge symmetry in these models. Thus, $Q_{B-L}(D_{45})=-2/3$ and
$Q_{B-L}(\bar D_{45})=2/3$.
{}From the above quantum numbers we see that \D and \bD are
actually leptoquarks [\LQ]. \D and \bD are superfields and therefore there are
two scalar and two fermionic leptoquarks in this model.
Similar states have been identified in Calabi--Yau [\GSW] and flipped $SU(5)
\times U(1)$ string models [\FLQ] and were called vector--like down quarks.

The phenomenology of \D and \bD including FCNC and baryon number ($B$)
violating
effects was investigated recently [\LQ]. It was shown that FCNC constraints are
easily satisfied due to the relatively large (i.e. $>TeV$) leptoquark masses
and very small (i.e. $<10^{-3}$) leptoquark Yukawa couplings. On the other
hand,
$B$ violating effects may be dangerous since \D and \bD may couple to
diquarks and and lepton--quark pairs simultaneously. These induce large $B$
violating operators unless some assumptions on the vanishing VEVs are made.

Leptoquark (from now on by leptoquarks or \D and \bD, we mean only the
fermionic ones since only they are relevant for our purposes) masses were
discussed in detail in Ref. [\LQ].
In general, one expects that \D  and \bD  get large masses (of
$O(10^{17}~GeV)$)
at the level of the cubic superpotential. Even if this is not the case \D and
\bD can get large masses from higher order (i.e. $N>3$) terms in the
superpotential and decouple from the low--energy spectrum. In Ref. [\LQ] it was
shown that all contributions to leptoquark masses (at $N=3$ and $N=5$) vanish
due to the cubic level F constraints which must be imposed to preserve
supersymmetry at $M_{Pl}$.

When hidden sector states are taken into account, there the are $N=6$ terms
$D_{45} \bar D_{45} T_2 \bar T_2 \Phi_{45} \Phi_2^+ (\xi_1+\xi_3)$
which may give large masses to \D and \bD. (Here $T_2, \bar
T_2$ are $5,\bar 5$ of the hidden $SU(5)_H$ gauge group.)
If $\l \Phi_2^+ \r \not=0$, then generically
$\l \Phi_2^+ \r \sim M/10 \sim 10^{17}~GeV$ and $\l T_2 \bar T_2 \r \sim
\Lambda_H^2$ where $\Lambda_H \sim 10^{14}~GeV$ is the hidden $SU(5)_H$
condensation scale [\CKM]. This gives $M_{D,\bar D} \sim 10^8~GeV$.
If $\l \Phi_2^+ \r=0$, then the \D \bD mass terms come
from the SUSY breaking VEVs. The VEVs vanishing due to
SUSY can become nonzero (and up to the $TeV$ scale) once SUSY is broken.
Therefore, when SUSY is broken, \D and \bD get $TeV$ scale masses from the
cubic superpotential, i.e. from the term
$W_{D,\bar D}=D_{45} \bar D_{45} \xi_3$
(since now $\l \xi_3 \r$ which vanished due to the supersymmetric F constraints
is $\sim O(TeV)$). Thus, in this model, there are two fermionic
leptoquarks with masses between $10^3~GeV$ and $10^8~GeV$ depending on the
scalar VEVs.

The leptoquarks, \D and \bD, may mix with down--like quarks. In fact,
there are nonrenormalizable terms which induce leptoquark mixing with
right--handed down quarks of the form
$$\eqalignno{
&d_3 D_{45} N_3 \Phi_{13} \Phi_3^+ \xi_i, &(1a) \cr
&d_2 D_{45} N_2 \Phi_2^- \xi_i, &(1b) \cr
&d_1 D_{45} N_1 \Phi_1^+ \xi_i, &(1c)}$$
where $\xi_i$ means $\xi_1+\xi_2$.
Similar mixing terms may also appear at higher
orders but we neglect them since they are suppressed relative to those given
above. $\l N_i \r $ appears in
nonrenormalizable terms which induce dimension four $B$ and lepton
number ($L$) violating operators [\NRT].
Explicitly, the $N>3$ terms which induce $B$ and $L$ violating terms
in the superpotential are [\NRT]
$$\eqalignno{&(u_3d_3+Q_3L_3)d_2N_2 \Phi_{45} \bar \Phi_2^-, &(2a) \cr
             &(u_3d_3+Q_3L_3)d_1N_1 \Phi_{45} \Phi_1^+, &(2b) \cr
             &u_3d_2d_2N_3 \Phi_{45} \bar \Phi_2^- +u_3d_1d_1N_3 \Phi_{45}
\Phi_1^+, &(2c) \cr
             &Q_3L_1d_3N_1 \Phi_{45} \Phi_3^+ +Q_3L_1d_1N_3 \Phi_{45} \Phi_3^+,
&(2d) \cr
             &Q_3L_2d_3N_2 \Phi_{45} \bar \Phi_3^- +Q_3L_2d_2N_3 \Phi_{45}
\bar \Phi_3^-, &(2e) }$$

In order to satisfy the constraints from the proton lifetime, the coefficients
of the above $B$ and $L$ violating operators must be $<10^{-13}$ (for
sparticles with masses of $O(TeV)$) [\BAR]. From this we get the constraint on
the sneutrino VEVs, $\l N_i \r \sim O(10^7~GeV)$ at most. There are no other
phenomenological constraints on $\l N_i \r$, therefore we conclude that
$0 \le \l N_i \r <10^7~GeV$.

In addition, there are leptoquark mixing terms with left--handed down quarks
such as
$$\eqalignno{
&Q_3 \bar D_{45} h_{45} H_{13} H_{23} V_3 \Phi_{45} \xi_i, &(3a) \cr
&Q_2 \bar D_{45} h_{45} H_{13} H_{23} V_2 \Phi_{45} \xi_i, &(3b) \cr
&Q_1 \bar D_{45} h_{45} H_{13} H_{23} V_1 \Phi_{45}.  &(3c) }$$
The only problem with these terms is the fact that supersymmetric F constraints
at the cubic level of the superpotential require $\l H_{13} \r=0$ at the
Planck scale (whereas $H_{23}$ may get a nonvanishing VEV) [\UP]. On the other
hand, it is plausible that higher order corrections
to the superpotential modify the F constraints in such a way as to allow a
large VEV for $H_{13}$. In the following we will assume this to be the case.

If the mixing terms in Eqs. (1) and (3) are nonzero, then we have a $4 \times
4$
down quark mass matrix, $M_d$ of the form (in the basis ($d,s,b,D_{45}$))
$$M_d=\left( \matrix{m_d&0&0& m_3^{\prime} \cr
                      0&m_s&0& m_2^{\prime} \cr
                      0&0&m_b& m_1^{\prime} \cr
                      m_3&m_2&m_1&M_D} \right) \eqno(4)$$
where we assume that there are no direct quark mixing terms. (This can be
easily
achieved by choosing the VEVs of $\bar V_i$ to be zero since direct mixing
terms
are proportional to $\l V_i \bar V_j \r$. [\CAB,\CKM]) The case with only
direct quark mixing terms was investigated previously [\CAB,\CKM]. It was
found that, with a proper choice of scalar VEVs (F and D flat direction) a
realistic CKM matrix can be obtained.
We also take the
up quark mass matrix, $M_u$, to be diagonal:
$$M_u=\left( \matrix{m_u&0&0 \cr
                      0&m_c&0 \cr
                      0&0&m_t} \right) \eqno(5)$$
There are no direct quark mixing terms in $M_u$ either. In any case, it can be
shown that the off--diagonal terms in $M_u$ do not affect the CKM matrix by
much. The CKM matrix arises mainly from the off--diagonal terms of $M_d$ [\CAB,
\CKM].
$m_i$ and $m_i^{\prime}$ which parametrize the leptoquark mixings with the
right and left--handed down quarks respectively, are obtained directly from the
mixing terms given by Eqs. (1) and (3).

The top mass is obtained at the cubic level from $u_1 Q_1 \bar h_1$ whereas
the bottom, charm and strange masses are obtained, as usual, from $N=5$ terms
[\NRT]
$$\eqalignno{&u_2 Q_2 (\bar h_{45} \Phi_{45} \bar \Phi_{23}+\bar h_1 \bar
\Phi_i^+ \bar \Phi_i^-), &(6a)\cr
&d_1 Q_1 h_{45} \Phi_1^+ \xi_2, &(6b) \cr
&d_2 Q_2 h_{45} \bar \Phi_2^- \xi_1. &(6c) }$$
The up and down quarks get masses from higher order terms [\UP]. Correct order
of magnitude masses for all quarks can be obtained by a proper choice of scalar
VEVs.

We now analyze the CKM matrix that arises from the above $M_d$ under different
assumptions about the mixing terms which are parametrized by $m_i$ and
$m_i^{\prime}$.

\bigskip
\centerline{\bf 3. The CKM matrix}

The left--handed down quark mixing matrix which arises from the above $M_d$ has
been obtained before. To first order in the small parameters $m_i/M_D$ and
$m_i^{\prime}/M_D$ it is given by [\DA]
$$V=\left( \matrix{1&\mu_{32}/m_s&\mu_{31}/m_b&-m_3^{\prime}/M_D \cr
                   -\mu_{32}^*/m_s&1&\alpha &-m_2^{\prime}/M_D \cr
                   -\mu_{31}^*/m_b&-\alpha^*&1&-m_2^{\prime}/M_D \cr
                   m_3^{\prime *}/M_D&m_2^{\prime *}/M_D&m_1^{\prime *}/M_D&
1} \right) \eqno(7)$$
where $\mu_{ij}=m_i^{\prime}m_j/M_D$ and $\alpha=[m_b/(m_b^2-m_s^2)][\mu_{21}
+(m_s/m_b)\mu_{12}^*]$. The terms of $V$ are at most linear in
the small parameters $\mu, m_i/M_D, m_i^{\prime}/M_D$. As we will see later,
higher order corrections to the elements of $V$ are negligible unless
these elments vanish.
The right--handed down quark mixing matrix is given by $U=V(m_i
\leftrightarrow m_i^{\prime *})$. From Eq. (7) we see that $|V|$ is symmetric
(up to corrections which will be discussed in the following) which is the most
important prediction of this scenario for quark mixing.

The CKM matrix is the $3 \times 3$ block of $V$ given above since there is no
mixing in the up quark sector (or $M_u$).
We want to see whether the elements of $V$ can give us realistic quark mixing
angles (we neglect all phases for our purposes or consider $|V|$).
Now, the $3 \times 3$ CKM matrix becomes nonunitary
because of the nonzero $4j$ and $i4$ elements in the $4 \times 4$ down quark
mixing matrix, $V$. The strongest bounds on the magnitude of the
new mixing terms parametrized by $m_i$ and $m_i^{\prime}$ arise from
the unitarity of the ($3 \times 3$) CKM matrix $V$ which
imposes $|V_{uD}|<0.07$ [\PDG] and from flavor changing $Z$ currents [\NIR]
which imposes $|Re V_{id}^* V_{is}|<2.4 \times 10^{-5}$, $i=u,c,t$.
We would like to obtain realistic quark mixing without violating these
phenomenological constraints for some choice of scalar VEVs (which appear in
Eqs. (1) and (3)).

{}From the mixing matrix $V$ given above, we see that the scalar VEVs must be
such that $\mu_{32}/m_s \sim 0.2$, $\mu_{31}/m_b \sim 10^{-3}$ and
$\alpha \sim 0.03$ to get the experimentally measured quark mixings.
These values can be obtained by a suitable choice of the elements of $M_d$
as follows. If $m_2/M_D \sim 1/5$, $m_3^{\prime} \sim m_s $ and $m_1/m_b
\sim 10$, then $V_{us}$ and $V_{ub}$ are of the correct order of magnitude.
If in additon, we also have $m_2^{\prime}/M_D \sim 3 \times 10^{-3}$ then
we also get a
realistic $V_{cb}$. We find that the unitarity constraint, $|V_{uD}|=
m_3^{\prime}/M_D<0.07$ is easily satisfied by the above choice of values.
In order to satisfy the constraint from FCNC, we must also satisfy
$|Re V_{id}^* V_{is}|=m_1^{\prime}m_2^{\prime}m_3^2/M_D^2m_b^2<2.4 \times
10^{-5}$.
This too can be easily satisfied by choosing $m_3/m_b \sim 2$ and
$m_1^{\prime}/M_D \sim 10^{-4}$.

Of course, all elements of $M_d$ that we chose above result from some choice
of scalar VEVs as dictated by the mixing terms given in Eqs. (1) and (3).
In the following, we take $M_D \sim 1~TeV$ by imposing $\l \Phi_2^+\r=0$.
($1~TeV<<M_D<10^5~TeV$ which is possible gives very small $m_i/M_D$ and
$m_i^{\prime}/M_D$ which in turn cannot produce appreciable quark mixing.)
A choice of scalar VEVs which will produce the desired values for
$m_i^{\prime}$ is
$$\l H_{13}, H_{23} \r \sim M \qquad \hbox{and} \qquad \l V_2, \Phi_{45},
\xi_1, \xi_2 \r \sim {M \over 10} \eqno(8)$$
with $\l V_1 \r \sim \l V_3 \r$ and $\l V_2 \r \sim 20 \l V_3 \r$.
We take $\l h_{45} \r \sim 150~GeV$ for our order of magnitude estimates.
On the other hand, in order to obtain the desired values for $m_i$ we can
choose
$$\l \Phi_{13}, \Phi_1^+, \Phi_2^-, \Phi_3^+ \r \sim {M\over 10} \eqno(9)$$
and $2\l N_1 \r=\l N_3 \r \sim TeV$ and $\l N_2 \r \sim 20~TeV$.

There are a large number of F and D flat solutions (sets of scalar VEVs) which
give the desired $m_i$ and $m_i^{\prime}$ for this scenario. In general the
F and D constraints will force other VEVs not required by the above mechanism
to be nonzero too. For example, the D constraint for the hidden $SU(5)$
requires at least one $\bar V_i$ with nonzero VEV (which together with the
$V_i$ may produce direct quark mixing).
Note that the required choice of scalar VEVs for this scenario is not the most
natural one even though it is certainly possible. Generically scalar VEVs in
these models are $\sim M/10$ whereas above we require relatively large VEVs
(of $O(M)$) for $H_{13}, H_{23}$ and a small VEV (of $O(M/10^2)$) for $V_2$.

{}From the mixing matrix $V$ in Eq. (7) we see that all the off--diagonal
elements
vanish if either all $m_i$ or $m_i^{\prime}$ are zero. This can happen if, for
example, either $\l N_i \r=0$ or $\l V_i \r=0$. In that case one has to go to
higher orders in the small parameters $\mu,m_i/M_D,m_i^{\prime}/M_D$ to obtain
the nonzero quark mixings. Higher order corrections to $V$ when $m_i=0$ are
given by [\DCC]
$$\eqalignno{&V_{ij}=\delta_{ij}+(1-\delta_{ij}){m_i^{\prime}m_j^{\prime}
\over M_D^2}{m_i^2 \over {(-1)^{\delta_{ij}} m_j^2-m_i^2}} &(10a) \cr
&V_{i4}=-V_{4i}=-{m_i^{\prime} \over M_D}, \qquad V_{44} \sim 1 &(10b,c)}$$
where we neglected all phases and $m_i$ in the above formulas are now the
down quark masses. This case is particularly interesting because it realizes
the Nelson--Barr mechanism [\NB] for the solution of the strong CP problem.
When all $m_i$ in $M_d$ given by Eq. (1) are zero, $Det(M_d)$ is real even
if the $m_i^{\prime}$ (but not the diagonal elements, i.e. the quark masses)
carry phases. Then if $\theta_{QCD}=0$ for some reason, the $\theta$ angle
does not get an additional contribution from the quark mass matrices since
$\theta_{quark}=argDet(M_uM_d)=0$. Due to the very small ratio $m_i^{\prime}
/M_D \sim 10^{-3-4}$ this scenario does not have any problems from
supersymmetric
processes either [\SNB]. The hope is to obtain large enough weak
CP violation (or quark mixing) solely from the phases of $m_i^{\prime}$ in Eq.
(4). From the mixing terms in Eq. (10) for this case we find that this hope
cannot be realized as was also noticed in Ref. [\SNB]. For example,
the Cabibbo mixing, $V_{32} \sim 0.2$
can only be obtained from Eq. (10a) if $m_i^{\prime}m_j^{\prime}/M_D^2 \sim 4$
which means that we need $m_i^{\prime}>M_D$ which is beyond the validity of
our approximations. (We remind that all of the above formulas are expansions
in the small parameters $\mu, m_i/M_D, m_i^{\prime}/M_D<<1$). This result was
also checked numerically and it was found that when $m_i^{\prime}>M_D$ higher
order corrections completely change the first order results.

The terms in Eq. (10) also give the first corrections to the elements of the
mixing matrix $V$. Since $m_i^{\prime}/M_D \sim 10^{-3-4}$ for our scalar VEVs,
these corrections are at most about $10^{-3}$ times the matrix elements and
therefore negligible compared to the lowest order terms in Eq. (7). From Eq.
(10a) we see that, contrary to the elements of $|V|$, these small corrections
are not symmetric. As a result, this scenario can only accomodate a CKM matrix
whose absolute value is symmetric up to a few parts in a thousand.

\bigskip
\centerline{\bf 4. Conclusions}

To summarize, we have obtained correct order of magnitude quark mixing solely
from the leptoquark--down quark mixing terms in the particular standard--like
superstring model examined.
There are a large number of F and D flat directions which produce this result
(among the infinitely many which do not). The leptoquark masses $M_{D,\bar D}$
cannot be much larger than a few $TeV$ for this scenario to work since
otherwise $m_i/M_D,m_i^{\prime}/M_D$ are too small to give appreciable quark
mixing. Fortunately, such low masses are possible in this model.
We found that when $\l N_i \r=0$ so that $m_i=0$ in $M_d$ one cannot get
large enough quark mixing (or weak CP violation)
and therefore the related Nelson--Barr mechanism is not realistic in this case.

The most important prediction of the scenario we discussed above is the
symmetry of $|V|$, the absolute value of the mixing matrix. This can be
easily seen from Eq. (7) for $V$ (up to the first corrections given by Eq. (10)
which are not symmetric). Since the first corrections to $V$ are at most
about a few parts in a thousand, it
will be very difficult to accomodate a nonsymmetric CKM matrix in this
scenario.
This property is model independent since it is a direct result of the form of
$M_d$ in Eq. (4) and does not depend on how $m_i$ and $m_i^{\prime}$ arise.

In addition, there are model dependent predictions. For example, from
Eqs. (1), (3) and (7) we find that
$${V_{ub} \over V_{cb}}={\l V_3 \r \over \l V_2 \r} \eqno(11)$$
up to a few percent. (The corrections to this equality which are about a few
percent mainly come from the approximation we use for $\alpha \sim
{m_2^{\prime}m_1/ m_b}$.) It would be intereseting to constrain the values
of $\l V_2,V_3 \r$ from other phenomenological phenomena and see if this
relation holds. This, however, is a model dependent prediction and will only
test this scenario in the framework of standard--like superstring models.

\bigskip
\centerline{\bf Acknowledgements}

This work was supported by the Department of Particle Physics and a Feinberg
Fellowship.

\vfill
\eject

\refout
\vfill
\eject

\end
\bye